\documentstyle[buckow]{article}

\def\beq{\begin{equation}}
\def\eeq{\end{equation}}

\begin {document}
\def\titleline{
Quantum mechanics from a Heisenberg-type equality
}
\def\authors{
Michael J. W. Hall\1ad and Marcel Reginatto\2ad}
\def\addresses{
\1ad Theoretical Physics, RSPSE \\
Australian National University\\
Canberra  ACT 0200   Australia.\\
{\tt michael.hall@anu.edu.au} \\
\2ad Physikalisch-Technische Bundesanstalt\\
Bundesallee 100\\
38116 Braunschweig   Germany
}
\def\abstracttext{
The usual Heisenberg uncertainty relation,
$\Delta X\Delta P\geq\hbar /2$, may be replaced by an exact equality
for suitably chosen measures of position and momentum uncertainty, 
which is valid for all wavefunctions.
This {\it exact} uncertainty relation, $\delta X\Delta P_{nc}\equiv\hbar /2$,
can be generalised to other pairs of conjugate observables such as
photon number and phase, and is sufficiently strong to provide the basis for
moving from classical mechanics to quantum mechanics.  In particular,
the assumption of a nonclassical momentum fluctuation, having a strength
which scales inversely with uncertainty in position, leads from the
classical equations of motion to the Schr\"{o}dinger equation.
}
\large
\makefront

\section{An exact uncertainty relation}
The well known Heisenberg inequality $\Delta X\Delta P\geq\hbar /2$ has
a fundamental significance for the interpretation of quantum theory - as
argued by Heisenberg himself, it
provides ``that measure of freedom from the limitations of classical
concepts which is necessary for a consistent description of atomic
processes'' \cite{WH}.
It might be asked whether this ``measure of freedom''
from classical concepts can be formulated
more precisely.  The answer is, surprisingly,
yes - and, as a consequence, 
the Heisenberg inequality can be replaced by an exact {\it equality},
valid for all pure states.

To obtain this equality, note that for a classical system,
the position and momentum observables can be measured 
simultaneously, to an arbitrary accuracy. For a quantum system we 
therefore define the {\it classical component} of the momentum 
to be that observable which is {\it closest} to the momentum
observable, under the constraint of being comeasurable with the position
of the system.  More formally, for the case that we have maximal
knowledge about the system, i.e., we know its wavefunction $\psi(x)$,
the classical component $P_{cl}^\psi$ of the momentum is
defined by the properties 
\beq \label{prop}
[X,P_{cl}^\psi]=0, \hspace{1.5cm} \langle\, (P-P_{cl}^\psi)^2\,\rangle_\psi =
{\rm minimum}.
\eeq

The unique solution to (\ref{prop}) is $P_{cl}^\psi=\int dx\,|x\rangle\langle
x|\,P_{cl}^\psi(x)$, where \cite{H1}
\beq \label{pcl}
P_{cl}^\psi(x) = \frac{\hbar}{2i}\left[\psi'(x)/\psi(x)-
\psi'^{*}(x)/\psi^*(x) \right] .
\eeq
Thus $P_{cl}^\psi(x)$ provides the best possible estimate of momentum for 
state $\psi(x)$ consistent with position measurement result $X=x$.  
It may also be recognised as the momentum current which appears in the
quantum continuity equation for the probability density $|\psi(x)|^2$,
as well as the momentum associated with the system in the de
Broglie-Bohm interpretation of quantum mechanics \cite{BB}.

Having a classical momentum component, it is natural to define the
corresponding {\it nonclassical} component of the momentum, 
$P_{nc}^\psi$, via the decomposition 
\beq \label{decomp}
P = P_{cl}^\psi + P_{nc}^\psi .
\eeq
The average of $P_{nc}^\psi$ is zero for state $\psi(x)$, and hence 
$P$ may be thought of as comprising a nonclassical fluctuation
about a classical average.  It is the nonclassical component which is
responsible for the commutation relation $[X,P]=i\hbar$. 
One has the related decomposition \cite{H1}
\beq \label{addvar}
(\Delta P)^2 = (\Delta P_{cl}^\psi)^2 + (\Delta P_{nc}^\psi)^2
\eeq
of the momentum variance into classical and nonclassical components,
and there is a similar decomposition of the kinetic energy.

The magnitude of the nonclassical momentum fluctuation, $\Delta P_{nc}^\psi$, 
provides a natural measure for that ``degree of freedom from the
limitations of classical concepts'' referred to by Heisenberg \cite{WH}.
Note that this
magnitude can be operationally determined from the statistics of $X$ and
$P$, via equations (\ref{pcl}) and (\ref{addvar}).  

It is remarkable that $\Delta P_{nc}^\psi$
satisfies an {\it exact} uncertainty relation \cite{H1}
\beq \label{ex}
\delta X \Delta P_{nc}^\psi \equiv \hbar/2,
\eeq
where $\delta X$ denotes a classical measure of position uncertainty,
called the ``Fisher length'', defined via \cite{RF}
\[ (\delta X)^{-2} = \int_{-\infty}^{\infty} dx\,\,p(x)\,
\left[(d/dx) \ln p(x)\right]^2
\]
for probability density $p(x)$.  There is thus a {\it precise} connection
between the  statistics of complementary observables.

The exact uncertainty relation (\ref{ex}) is far stronger than the usual 
Heisenberg inequality (the latter follows immediately via
(\ref{addvar}) and the Cramer-Rao inequality $\Delta X \geq \delta X$ of
classical statistical estimation theory). For example, 
suppose that at some time the wavefunction $\psi(x)$ is
confined to some interval.  Then, since $\ln p(x)$ 
changes from $-\infty$ to a finite value in any neighbourhood containing
an endpoint of the interval, the Fisher length $\delta X$ vanishes.  The
exact uncertainty relation (\ref{ex}) thus immediately implies that $\Delta
P_{nc}^\psi$, and hence $\Delta P$, is unbounded for any such confined
wavefunction.  
 
Exact uncertainty relations may be generalised and/or applied
to, for example, density operators, higher dimensions, energy bounds,
photon number and phase, and entanglement \cite{H1}.  A conjugate
relation, $\delta P \Delta X_{nc}^\psi\equiv\hbar/2$, may also be derived. 
In the following section the very {\it existence} of exact uncertainty
relations is used as a basis for deriving the Schr\"{o}dinger
equation.

\section{QM from an exact uncertainty principle}

Landau and Lifschitz wrote, referring to the Heisenberg uncertainty
principle, that ``this principle in itself does
not suffice as a basis on which to construct a new mechanics of
particles'' \cite{LL}.  However, the existence of exact uncertainty
relations for quantum systems raises anew the question of
whether the uncertainty principle, at the conceptual core of the standard
Copenhagen interpretation of quantum theory, can be put in a form strong
enough to provide an axiomatic means for moving from classical to
quantum equations of motion.  The corresponding {\it exact} uncertainty
principle would thus be on a par with alternative derivations based on
the principle of superposition, $C^*$-algebras, quantum logics, etc. 

It has recently been shown that an ``exact uncertainty principle'' does
indeed exist, where the Schr\"{o}dinger equation may be derived via the
postulate that classical systems are subjected to random momentum
fluctuations of a strength inversely proportional to uncertainty in
position \cite{HR}.  This new approach is summarised below (though using
a different method of proof than in \cite{HR}).

Now, in any axiomatic-type construction of quantum mechanics one must first 
choose a classical starting 
point, to be generalised or modified appropriately.  The
starting point here is a statistical one - the classical motion of an 
ensemble of particles - and indeed most of the assumptions to be made 
below will be seen to have a statistical character.  

Consider then a classical ensemble of $n$-dimensional particles of mass $m$ 
moving under a potential $V$. The motion may be described via the
Hamilton-Jacobi and continuity equations  
\[ \frac{\partial s}{\partial t} +
\frac{1}{2m}|\nabla s|^2 + V = 0, \hspace{1.5cm}
\frac{\partial p}{\partial t} +\nabla\cdot\left[ p\frac{\nabla s}{m}
\right] = 0,
\]
respectively, for the ``momentum potential'' $s$ 
and the position probability density
$p$.  These equations follow from the variational principle
$\delta L=0$ with Lagrangian
\beq \label{lc}
L = \int dt\,d^nx\, p \left[ \frac{\partial s}{\partial t} +
\frac{1}{2m}|\nabla s|^2 + V\right],
\eeq
under fixed endpoint variation with respect to $p$ and $s$. 
This Lagrangian is therefore chosen as our classical starting point.

It is now assumed that the classical Lagrangian (\ref{lc}) must be modified, due
to the existence of random momentum fluctations.
The nature of these
fluctuations is not important to the argument - they may be postulated
to model experimental evidence that the momentum variance is sometimes
greater than the expected $\int dx\,p\,|\nabla s-\langle\nabla s\rangle|^2$; 
or they may be regarded as a device to make the system irreducibly
statistical (since such fluctuations imply the velocity 
relation ${\bf v}=m^{-1}\nabla s$ no longer holds, and hence cannot be
integrated to give corresponding trajectories). The assumption is simply 
that {\it the momentum associated with position $x$ is given by}
\[ {\bf P} = \nabla s + {\bf N},\]
{\it where the fluctuation term ${\bf N}$ vanishes on the average at 
each point $x$}. 
The physical meaning of $s$ thus changes to being
an {\it average} momentum potential.

It follows that the average kinetic energy $\langle |\nabla s|^2\rangle/(2m)$ 
appearing in (\ref{lc})
should be replaced by $\langle |\nabla s+{\bf N}|^2\rangle/(2m)$ (where
$\langle\,\rangle$ now denotes the average over fluctuations {\it and}
position), giving rise to the modified Lagrangian 
\beq  \label{lmod}
L' = L + (2m)^{-1}\int dt\,\langle {\bf N}\cdot{\bf N}\rangle =
L + (2m)^{-1}\int dt\,(\Delta N)^2 ,
\eeq
where $\Delta N=\langle {\bf N}\cdot{\bf N}\rangle^{1/2}$ is a measure
of the strength of the fluctuations. 
Note that the additional term in (\ref{lmod})
corresponds to the kinetic energy of the fluctuations, and so is
positive.  

The additional term is specified uniquely, up to a multiplicative constant,
by the following three assumptions:

{\it (1) Action principle}:  $L'$ is a scalar
Lagrangian with respect to the fields $p$ and $s$, where the variational 
principle $\delta L'=0$ yields causal equations of motion. 
Thus
\[ (\Delta N)^2 = \int d^nx\,\,p\,f(p, \nabla p, \partial p/\partial t,
s, \nabla s, \partial s/\partial t, {\bf x}, t)\]
for some scalar function $f$. 

{\it (2) Additivity}: If the system comprises two independent 
non-interacting subsystems 1 and 2, with $p=p_1p_2$, then the Lagrangian
decomposes into additive subsystem contributions.  Thus
\[ f = f_1 + f_2 \hspace{1cm}{\rm for}\hspace{1cm}p=p_1p_2, \]
where subscripts denote corresponding subsystem quantities. Note this
is equivalent to the statistical assumption that the corresponding
subsystem fluctuations are linearly uncorrelated, i.e., $\langle {\bf N_1\cdot
N_2}\rangle=0$.

{\it (3) Exact uncertainty principle}: The strength of the momentum
fluctuation at any given time is determined by, and scales inversely with, 
the uncertainty in position at that time.  Thus
\[ \Delta N\rightarrow k\Delta N \hspace{1cm}{\rm for}\hspace{1cm}
{\bf x}\rightarrow {\bf x}/k,\]
and moreover, since position uncertainty is entirely characterised by the
probability density $p$ at a given time, 
the function $f$ cannot depend on $s$, nor
explicitly on $t$, nor on the time-derivative of $p$.

We now have the following theorem \cite{HR}:

{\bf Theorem:}  {\it The above three assumptions of an action principle,
additivity, and an exact uncertainty principle imply that} 
\[ (\Delta N)^2 = C\int d^nx\,\,p\,|\nabla\ln p|^2,\]
{\it where $C$ is a positive universal constant.}

A (new) proof of the above theorem is outlined below.  Here its main
consequence is noted.  In particular, it follows directly via (\ref{lmod}) that
the equations of motion for $p$ and $s$,
corresponding to the variational principle $\delta
L'=0$, can be expressed as the single complex equation 
\[ i\hbar\frac{\partial\psi}{\partial t}= -\frac{\hbar^2}{2m}\nabla^2\psi
+ V\psi,\]
where one defines $\hbar:=2\sqrt{C}$ and $\psi:=\sqrt{p}e^{is/\hbar}$.  Thus
the above postulates yield equations of motion equivalent to the
Schr\"{o}dinger equation. It can be shown that the mapping from
fields $p$ and $s$ to the wavefunction $\psi$ arises naturally 
from seeking canonical transformations which map  
to ``normal modes'' of the system \cite{HR}. 
It is remarkable that a linear equation results from
the assumptions used. 

\section{Proof of theorem} 

To see how the above theorem follows, note first that the action
principle and exact uncertainty principle imply that the scalar function
$f$ depends only on $p$, $\nabla p$, and ${\bf x}$.  
It can therefore be written in the form
\[ f=g(u,v,w,r^2),\]
where
\[ u=\ln p,\hspace{0.5cm} v=({\bf x}\cdot\nabla p)/p,\hspace{0.5cm} w=|\nabla
p|^2/p^2,\hspace{0.5cm}r^2={\bf x\cdot x} .\]

For $p=p_1p_2$ one finds $u=u_1+u_2$, $v=v_1+v_2$, and $w=w_1+w_2$, and
hence the additivity assumption implies that 
\[ g(u_1+u_2,v_1+v_2, w_1+w_2, r_1^2+r_2^2)=g_1(u_1,v_1,w_1,r_1^2)
+g_2(u_2,v_2,w_2,r_2^2).\]
Thus $g$ must be linear in $u$, $v$, $w$ and $r^2$, yielding 
\beq \label{f}
f = A\ln p + B\frac{{\bf x\cdot}\nabla p}{p}+C\frac{|\nabla p|^2}{p^2}+D{\bf
x\cdot x},
\eeq 
where $A$, $B$, $C$ and $D$ are universal constants. Note that the first
term corresponds to an entropic potential in the Lagrangian $L'$. 

Finally, noting that $p(x)\rightarrow k^np(kx)$ under the transformation
${\bf x}\rightarrow {\bf x}/k$,
the exact uncertainty principle forces $A=B=D=0$, and the theorem is
proved (where the positivity of $C$ follows from the positivity of
$(\Delta N)^2$).

\section{Conclusions}

It has been shown that the Heisenberg uncertainty relation may be
upgraded to an exact uncertainty relation, and that 
a corresponding {\it exact} uncertainty principle may be
used as the single nonclassical element necessary for obtaining the
Schr\"{o}dinger equation.  Thus the uncertainty principle is able to 
provide not
only a conceptual underpinning of quantum mechanics, but an axiomatic
underpinning as well.  The above approach immediately generalises to
include electromagnetic potentials, and work on  
generalisations to systems with
spin and to quantum fields is in progress. 

The form chosen for the exact uncertainty principle, that classical
systems are subject to random momentum fluctuations of a strength
inversely proportional to uncertainty in position, is of course motivated by the
momentum decomposition (\ref{decomp}) and exact uncertainty relation (\ref{ex})
holding for quantum systems.  Thus, not
surprisingly, there are a number of connections between the latter uncertainty
relation and the Theorem of section 2.
For example, for a 1-dimensional system, the Theorem immediately implies the
uncertainty relation
\[\delta X \Delta N =\sqrt{C}=\hbar/2\]
for the momentum fluctations, which may be compared to (\ref{ex}). 

The exact uncertainty principle has a type of ``nonlocality'' built into
it: the form of $\Delta N$ specified by the Theorem 
implies that a change in the 
position probability density arising from actions on one subsystem (eg, a
position measurement),  will typically
influence the behaviour of a second subsystem correlated with the first.  This
nonlocality corresponds to quantum entanglement, and has been analysed
to some extent via exact uncertainty relations in \cite{H1}.

It is worth noting that the approach here, based on exact
uncertainty, is rather different from other approaches which assign
physical meaning to fields $p$
and $s$ related to the wavefunction.  
For example, in the de Broglie-Bohm approach \cite{BB}, there are {\it no}
momentum fluctuations, and the classical equations of motion for $p$ and
$s$ are instead modified by adding a mass-dependent ``quantum potential'',
$Q$, to the classical potential term in the Hamilton-Jacobi equation. 
The form of this quantum potential is left unexplained, and is
interpreted as
arising from the influence of an associated wave acting on the system.
In contrast, in the exact uncertainty approach $\nabla s$ is an
{\it average} momentum, the form 
of an additional kinetic energy term arising from random momentum
fluctuations is {\it derived}, and no associated wave is assumed.  The 
formal connection between the two approaches is the relation 
\[ \delta(L'-L) = \int dt\,d^nx\,\,Q \,\delta p.\]

Finally, the exact uncertainty approach is also very different from the
stochastic mechanics approach \cite{NE}.  The latter postulates the
existence of a classical stochastic process in configuration space, with a drift
velocity assumed to be the gradient of some scalar, and defines 
an associated time-symmetric 
``mean acceleration'' ${\bf a}$ in terms of averages over both
the stochastic process
and a corresponding time-reversed process, which is postulated
to obey Newton's law
$m{\bf a}=-\nabla V$. In contrast, the exact uncertainty 
approach does not rely on a
classical model of fluctuations, nor on a new definition of acceleration, 
nor on properties of stochastic processes running backwards in time. The
formal connections between the approaches are 
\[ \nabla s=m{\bf u}, \hspace{1.5cm}(\Delta N)^2 = m^2\langle{\bf v\cdot v}
\rangle,\]
where ${\bf u}+{\bf v}$ and ${\bf u}-{\bf v}$ are the drift
velocities of the forward-in-time and backward-in-time processes respectively.  
It should be noted that $\langle {\bf u\cdot v}\rangle\neq 0$, and
hence one cannot identify $m{\bf v}$ with the random momentum fluctuation 
${\bf N}$.



\begin{thebibliography}{77}

\bibitem{WH}W. Heisenberg,{\it The Physical Principles of the Quantum
Theory} (Dover, USA, 1930), page 4.

\bibitem{H1} M.J.W. Hall, Phys. Rev. A {\bf 64} (2001) 052103
[quant-ph/0107149].

\bibitem{BB} D. Bohm, Phys. Rev. {\bf 85} (1952) 143, 187.

\bibitem{RF} R.A. Fisher, Proc. Cambridge Philos. Soc. {\bf 22} (1925)
700.

\bibitem{LL} L.D. Landau and E.M. Lifschitz, {\it Quantum Mechanics},
3rd edition (Pergamon, Oxford, 1977), page 2.

\bibitem{HR} M.J.W. Hall and M. Reginatto [quant-ph/0102069].

\bibitem{NE} E. Nelson, Phys. Rev. {\bf 150} (1966) 1079; E. Nelson, {\it Dynamical Theories of Brownian Motion}
(Princeton University Press, USA, 1967).
\end{thebibliography}
\end{document}